\documentclass[11pt,preprint]{aastex}
\usepackage{epsfig}
\newcommand       \Angstrom     {\,{\rm \AA}} 
          
\newcommand       \cm           {\,{\rm cm}}

\newcommand       \erg          {\,{\rm erg}}

\newcommand	  \g		{\,{\rm g}}
\newcommand       \K            {\,{\rm K}}

\newcommand	  \kpc		{\,{\rm kpc}}
\newcommand	  \s		{\,{\rm s}}

\newcommand       \Qabs	        {Q_{\rm abs}}
\newcommand       \Qabsuv	{Q_{\rm abs}^{\rm UV}}

\newcommand       \kabs         {\kappa_{\rm abs}} 
\newcommand       \kabsuv       {\kappa_{\rm abs}^{\rm UV}} 
\newcommand       \simlt        {\lesssim}
\newcommand       \simgt        {\gtrsim}

\newcommand       \mum          {\,{\rm \mu m}}

\newcommand	  \Teff	        {T_{\rm eff}}
\newcommand       \msun         {\,{m_\odot}}

\newcommand       \simali       {\sim\,}
\newcommand       \rmin         {r_{\rm min}}
\newcommand       \rmax         {r_{\rm max}}
\newcommand       \Eabs         {E^{\rm tot}_{\rm abs}\left(\rm TiC\right)}
\newcommand       \Eemsn        {E^{\rm tot}_{\rm em}\left(\rm 21\mum\right)}
\newcommand       \mtic         {m_{\rm TiC}^{\rm tot}}
\newcommand       \mticmin      {m_{\rm TiC}^{\rm min}}
\newcommand       \mticmax      {m_{\rm TiC}^{\rm max}}
\newcommand       \rhotic       {\rho_{\rm TiC}}
\newcommand       \rstar        {r_\star}
\newcommand       \Fstar        {F_{\lambda}^{\star}}
\newcommand       \HD           {{\rm HD\,56126}}
\newcommand       \menv         {m_{\rm env}}

\newcommand	  \cexttot	{C_{\rm ext}^{\rm tot}}
\newcommand	  \cabstot	{C_{\rm abs}^{\rm tot}}
\newcommand	  \cscatot	{C_{\rm sca}^{\rm tot}}
\newcommand	  \Vtot	        {V_{\rm tot}}
\newcommand	  \mtickk	{m_{\rm TiC}^{\rm KK}}
\newcommand	  \magni	{\,{\rm mag}}
\newcommand{\figwidth}{4.0in}
\countdef\decade=200
\advance\decade by \year
\countdef\hours=201
\advance\hours by \time
\divide\hours by 60
\countdef\mins=202
\advance\mins by \hours
\multiply\mins by 60
\multiply\hours by 100
\countdef\miltime=203
\advance\miltime by \hours
\advance\miltime by \time
\advance\miltime by -\mins

\shorttitle{On TiC Nanoparticles in Post-AGB Stars}

\begin{document}

\title{
%------------- enable for labelling preprint ---------------------------
 \vspace*{-2.0em}
  {\normalsize\rm To appear in {\it The Astrophysical Journal Letters},
  the 2003-December-20th issue}\\
 \vspace*{1.0em}
%-----------------------------------------------------------------------
On Titanium Carbide Nanoparticles as the Origin of the 21 Micron Emission 
Feature in Post-Asymptotic Giant Branch Stars
%\\{\small DRAFT: \today ~~}
	 }
%\author{Aigen Li\altaffilmark{1,2}}
%\altaffiltext{1} {Theoretical Astrophysics Fellow}
%\altaffiltext{2} {Departments of Astronomy and Planetary Sciences,
%                  University of Arizona, Tucson, AZ 85721;
%                  {\sf agli@lpl.arizona.edu}}
\author{Aigen Li}
\affil{Theoretical Astrophysics Program,
       Lunar and Planetary Laboratory and Steward Observatory,
       %Departments of Astronomy and Planetary Sciences,
       University of Arizona, Tucson, AZ 85721;\\
        {\sf agli@lpl.arizona.edu}}

\begin{abstract}
Titanium carbide (TiC) nanocrystals were recently proposed 
as the carrier of the mysterious 21$\mum$ emission feature 
observed in post-asymptotic giant branch stars,
based on their close spectral match
and the presolar nature of meteoritic TiC nanograins
(which reveals their stellar ejecta origin).
But we show in this {\it Letter} that 
the Kramers-Kronig dispersion relations, which relate 
the wavelength-integrated extinction cross section to 
the total dust mass, would impose a lower bound on the TiC mass.
This Kramers-Kronig lower limit
exceeds the maximum available TiC mass by 
a factor of at least $\simali$50,
independent of the absolute value of 
the ultraviolet/visible absorptivity of nano TiC.
The TiC model is therefore readily
ruled out by the Kramers-Kronig physical principle.
%
%We examine the proposal of titanium carbide (TiC) 
%nanoparticles as a carrier of the mysterious 21$\mum$ 
%emission feature observed in post-asymptotic giant branch 
%(AGB) stars by comparing the maximum {\it available} with 
%the minimum {\it required} TiC dust mass inferred 
%from the nano-TiC model. While this model appears 
%promising because of the close agreement of 
%the laboratory absorption spectra of 
%nanometer-sized TiC clusters 
%with the observed 21$\mum$ emission feature, 
%the amount of TiC dust mass 
%required to explain the observed feature 
%is unclear due to the lack of knowledge of the ultraviolet
%(UV) and visible optical properties of nano TiC materials.  
%If the UV/visible absorption properties of TiC nanograins
%are like their bulk counterparts, the model-required 
%TiC dust mass would exceed the maximum available TiC mass
%by over two orders of magnitude. 
%One may argue that nano TiC might have a much higher
%UV/visible absorptivity 
%%and a wavelength dependence optimized to enhance 
%%their ability in absorbing stellar radiation 
%so that the available TiC mass may be sufficient 
%to account for the observed 21$\mum$ emission feature.
%However, the Kramers-Kronig dispersion relations which 
%relate the wavelength-integrated extinction cross sections 
%to the total dust mass, would impose a lower bound on 
%the TiC mass. This Kramers-Kronig lower limit exceeds 
%the maximum available TiC mass by a factor of $\simali$50. 
%Therefore, TiC nanoparticles are unlikely responsible for
%the post-AGB 21$\mum$ feature.
\end{abstract}

\section{Introduction\label{sec:intro}}
A prominent broad (with a FWHM of $\simali$2$\mum$)
emission feature at about 21$\mum$\footnote{%
  \label{ftnt:iso21um}
  High resolution spectra obtained with 
  the {\it Short-Wavelength Spectrometer} (SWS) instrument 
  (with a resolution of $\lambda/\Delta\lambda \approx 2000$)
  on board the {\it Infrared Space Observatory} show that 
  this feature actually peaks at $\simali$20.1$\mum$
  (Volk, Kwok, \& Hrivnak 1999). But for historical reasons,
  this feature continues to be referred to as the ``21$\mum$
  feature''. 
  }
was discovered by Kwok, Volk, \& Hrivnak (1989) in an analysis 
of the {\it Infrared Astronomical Satellite} (IRAS) 
{\it Low Resolution Spectrometer} (LRS)
%(LRS; with a resolution 
%of $\lambda/\Delta\lambda \approx 20$) 
spectra of four carbon-rich post-asymptotic giant branch (AGB) stars,
and subsequently confirmed by both ground-based
%(Kwok, Hrivnak, \& Geballe 1995; Justtanont et al.\ 1996),
and airborne {\it Kuiper Airborne Observatory} 
and ISO observations 
(see Kwok, Volk, \& Hrivnak 1999 and references therein).
%and airborne KAO ({\it Kuiper Airborne Observatory}; 
%Omont et al.\ 1995) and ISO (Volk et al.\ 1999) observations.
So far, this feature has been detected in 
twelve post-AGB stars (commonly termed as 
the ``21$\mum$ sources''; Kwok et al.\ 1999)\footnote{%
  A weak 21$\mum$ feature was recently reported for three 
  planetary nebulae (PNs; Hony, Waters, \& Tielens 2001;
  K. Volk 2003, in preparation).
  }, 
with little shape variation found between different sources. 
These stars have quite uniform properties: 
they are mostly metal-poor carbon-rich F and G supergiants 
with strong infrared (IR) excesses
and overabundant $s$-process elements
(see Kwok et al.\ 1999).

The carrier of this mysterious 21$\mum$ feature 
remains unidentified, although many candidate materials 
have been proposed including 
iron oxides Fe$_2$O$_3$ or Fe$_3$O$_4$, 
%such as maghemite Fe$_2$O$_3$ 
%or magnetite Fe$_3$O$_4$, %(Cox 1990), 
%carbonaceous materials such as 
hydrogenated amorphous carbon,
%(Buss et al.\ 1990; Grishko et al.\ 2001), 
hydrogenated fullerenes, %(Webster 1995), 
hydrogenated nanodiamond, 
%(Koike et al. 1995; Hill, Jones, \& d'Hendecourt 1998),
amides (thiourea or urea ${\rm OC\left[NH_2\right]_2}$), 
%(thiourea or urea ${\rm OC\left[NH_2\right]_2}$; 
%Sourisseau, Coddens, \& Papoular 1992), 
SiS$_2$, %(Goebel 1993; Begemann et al.\ 1996), 
oxygen-bearing side groups in coal 
(see Kwok et al.\ 1999 and references therein),
%(Papoular 2000),
and more recently
titanium carbide (TiC) nanoclusters (von Helden et al.\ 2000),
SiC (Speck \& Hofmeister 2003),
and stochastically-heated silicon core-SiO$_2$ mantle
nanograins (Smith \& Witt 2002; Li \& Draine 2002).

The nano-TiC model seems attractive.
While bulk TiC does not show any noticeable
feature near 20.1$\mum$ (Henning \& Mutschke 2000),
laboratory absorption spectra of TiC nanocrystals
containing 27--125 atoms exhibit a distinct feature 
at $\simali$20.1$\mum$, closely resembling the astronomical
21$\mum$ emission feature both in peak position, 
width, and in spectral details (von Helden et al.\ 2000).
This model further gains its strength from the identification
of presolar TiC grains (with radii $\simali$100$\Angstrom$) 
in primitive meteorites as inclusions embedded in 
micrometer-sized presolar graphite grains (Bernatowicz et al.\ 1996). 

Since Ti is a rare element, it is important to know 
whether the amount of TiC dust required to account for 
the observed 21$\mum$ feature is a reasonable quantity.
Since the optical properties of nano TiC in 
the ultraviolet (UV)/visible wavelength range are unknown,
one has to rely on those of bulk TiC as a starting point.
Such an attempt has recently been made for 
the post-AGB star $\HD$ by Hony et al.\ (2003).
They found that to explain the observed 21$\mum$ feature,
the TiC model requires a much higher UV/visible absorptivity
than that of bulk TiC (by a factor of $\simgt$20). 
However, we will demonstrate in this {\it Letter}
how the TiC absorption integrated over a finite wavelength range
places a lower limit on the TiC dust mass
through the Kramers-Kronig dispersion relations. 
For $\HD$, this Kramers-Kronig lower limit exceeds 
the maximum available TiC mass by a factor of $\simgt$50,
independent of the absolute value of the UV/visible absorptivity.
Therefore, increasing the UV/visible absorptivity
is unable to avoid the TiC abundance problem.

\section{HD 56126: A Test Case\label{sec:obs}}
%%$\HD$ ($\equiv$\,IRAS07134+1005), 
$\HD$, a bright (visual magnitude $\simali$8.3) 
post-AGB star with a spectral type of F0-5I,
is one of the four 21$\mum$ sources originally discovered 
by Kwok et al.\ (1989). Mid-IR imaging of this object 
at 11.9$\mum$ shows that its circumstellar dust is confined
to an area of 1.2$^{\prime\prime}$--2.6$^{\prime\prime}$ from
the star (Hony et al.\ 2003). Detailed modeling of its dust IR
spectral energy distribution suggested a $dn(r)/dr \sim 1/r$
dust spatial distribution at 
$\rmin \simlt r \simlt \rmax$, where $r$ is the
distance from the star, $\rmin$ and $\rmax$ are respectively
the inner and outer edge of the $\HD$ dusty envelope
(Hony et al.\ 2003). If TiC nanograins are indeed present
in $\HD$ and follow the distribution of the bulk 
(hydrogenated) amorphous carbon dust,
the total power absorbed by the TiC dust would be\footnote{%
  The circumstellar envelope around $\HD$ is assumed to 
  be optically thin. If there actually exists appreciable
  extinction (e.g., Hony et al.\ [2003] found an average
  visual extinction of $\simali$1.1$\magni$), the conclusion 
  of this {\it Letter} would be strengthened since when exposed 
  to an attenuated stellar radiation field, the TiC model 
  would require more TiC dust to account for the same 
  21$\mum$ feature strength.
  %than that for the optically thin case.  
  }
\begin{equation}\label{eq:Eabs}
\Eabs = \mtic \frac{\rstar^2}{2}
\frac{\ln\left(\rmax/\rmin\right)}{\rmax^2-\rmin^2}
\int_{912\Angstrom}^{\infty} \kabs(a,\lambda)\,\Fstar d\lambda ~~,
\end{equation}
where $\mtic$ is the total mass of 
the TiC nanoparticle component,
$\kabs(a,\lambda)$ is the mass absorption coefficient 
(${\rm cm^2}\g^{-1}$) for nano TiC grains of size $a$ 
at wavelength $\lambda$, $\rstar$ is the stellar radius,
$\Fstar$ is the flux per unit wavelength 
(${\rm erg}\s^{-1}\cm^{-2}\mum^{-1}$)
at the top of the illuminating star's atmosphere.
In the nanometer size domain, TiC grains are
in the Rayleigh limit and hence $\kabs(\lambda)$ is 
independent of size $a$ in the wavelength range where 
the stellar radiation peaks. Following Hony et al.\ (2003), 
we will adopt a distance of $d\approx 2.4\kpc$ to the star
(and therefore $\rmin\approx 4.3\times 10^{16}\cm$,
$\rmax\approx 9.3\times 10^{16}\cm$),
a stellar radius of $\rstar\approx 49.2\,r_\odot$
($r_\odot$ is the solar radius),
a stellar luminosity of $L_\star\approx 6054\,L_\odot$
($L_\odot$ is the solar luminosity),
and approximate the $\HD$ stellar radiation 
by the Kurucz (1979) model atmospheric spectrum 
with $\Teff = 7250\K$ and $\log g=1.0$.

The total power emitted in the 21$\mum$ feature 
of $\HD$ was estimated to be 
$\Eemsn \approx 1.0\times 10^{36}\erg\s^{-1}$ (Hony et al.\ 2003).
A {\it lower} limit on the total mass of the TiC dust required 
to account for the observed 21$\mum$ feature can be obtained 
by assuming that all the energy absorbed by the TiC dust 
would be emitted solely in this feature:
\begin{equation}\label{eq:mtic}
\mticmin = \Eemsn \frac{2}{\rstar^2}
\frac{\rmax^2-\rmin^2}
{\ln\left(\rmax/\rmin\right)}/\int_{912\Angstrom}^{\infty} 
\kabs(\lambda,a)\,\Fstar d\lambda ~~.
\end{equation}
Apparently, the required TiC dust mass $\mticmin$ is sensitive 
to the material's UV/visible absorptivity. 
%The resulting integrated total absorption cross sections are
%\begin{equation}\label{eq:cabstot}
%\int_{912\Angstrom}^{\infty}
%\cabstot(\lambda) d\lambda = \frac{3 \mticmin}{4\rhotic}
%\int_{912\Angstrom}^{\infty} \frac{\Qabs(\lambda)}{a}d\lambda ~~.
%\end{equation}

On the other hand, an {\it upper} limit on the total mass 
of the TiC dust which could be present in the circumstellar 
envelope around $\HD$ can be estimated 
from its photospheric Ti abundance 
${\rm Ti/H \approx 1.3\times 10^{-8}}$
(van Winckel \& Reyniers 2000) 
and the envelope mass (including both gas and dust) 
$0.16\simlt \menv \simlt 0.44\msun$ (Hony et al.\ 2003): 
$\mticmax \approx 2.5\times 10^{-7}\msun$.

\begin{figure}[ht]
\begin{center}
\epsfig{
        %file=f1.eps, %for ApJL submission.
        file=f1.cps,
        width=\figwidth,angle=0}
\end{center}\vspace*{-1em}
\caption{
        \label{fig:qabs}
        \footnotesize
        Mass absorption coefficients $\kabs(\lambda)$ 
        of TiC nanograins of radii $a=10\Angstrom$
        calculated (1) from the optical constants of 
        bulk TiC (solid line),
        (2) from the Hony et al.\ (2003) hypothetical formula
        with $\kabsuv = 5.6\times 10^{6}\cm^2\g^{-1}$
        (see Eq.[\ref{eq:kabshony}]; dashed line),
        or (3) from a hypothetical Drude profile 
        with $\kabsuv = 9.6\times 10^{6}\cm^2\g^{-1}$
        (see \S\ref{sec:kk}; dot-dashed line). 
        The IR part is the laboratory measured 21$\mum$ spectrum
        of TiC nanocrystals (von Helden et al.\ 2000) 
        approximated by a Lorentz profile (Chigai et al.\ 2003).
        It is apparent that the Hony et al.\ (2003) and Drude
        hypothetical absorption spectra greatly enhance 
        the ``ability'' of TiC nanograins in absorbing
        the $\HD$ starlight.
        }
\end{figure}

\section{Ultraviolet/Visible Absorption Properties 
of TiC Nanograins\label{sec:qabs}}
Since the UV/visible absorption properties of nano TiC
are unknown, we will first represent them by those of bulk TiC. 
Using Mie theory and the optical constants measured for 
bulk TiC samples (Koide et al.\ 1993), 
we calculate the UV/visible absorption efficiency $\Qabs$
of spherical TiC nanograins.
For illustration, we plot in Figure \ref{fig:qabs}
the mass absorption coefficients 
($\kabs = 3\Qabs/\left[4 a \rhotic\right]$
where $\rhotic \approx 4.92\g\cm^{-3}$ 
is the TiC mass density) of TiC spheres of
radii $a$=10$\Angstrom$ calculated from
the optical constants of bulk TiC. 
Using these $\kabs$ values, we derive a lower limit 
of $\mticmin\approx 9.3\times10^{-5}\msun$ on
the TiC dust mass from Eq.(\ref{eq:mtic}).
Apparently, if nano TiC grains do not have a much higher
UV/visible absorptivity than their bulk counterparts,
they can not be the 21$\mum$ feature carrier since the minimum
required TiC dust mass is larger than the maximum available
TiC mass ($\mticmax \approx 2.5\times 10^{-7}\msun$; 
see \S\ref{sec:obs}) by over two orders of magnitude!  

Hony et al.\ (2003) assumed a hypothetical 
absorption efficiency for TiC nanograins
\begin{equation}\label{eq:kabshony}
\kabs(\lambda) = 
\left\{\begin{array}{lr} 
\kabsuv ~, ~~912\Angstrom \le \lambda \le 1\mum ~;\\ 
\kabsuv\left(2-\lambda\right) ~, ~~1\mum < \lambda \le 2\mum ~;\\
0 ~, ~~ \lambda > 2\mum ~;
\end{array}\right.
\end{equation}
where $\kabsuv$ is the ``constant level'' UV/visible 
mass absorption coefficient.
They concluded that the astronomical 21$\mum$ feature
could {\it only} be accounted for by the TiC model with 
a reasonable amount of TiC dust mass {\it if} nano TiC grains have 
an UV/visible absorption efficiency as high as
$\Qabsuv\approx 8$ 
where $\Qabsuv = \left(4/3\right) a \rhotic \kabsuv$.

\begin{figure}[ht]
\begin{center}
\epsfig{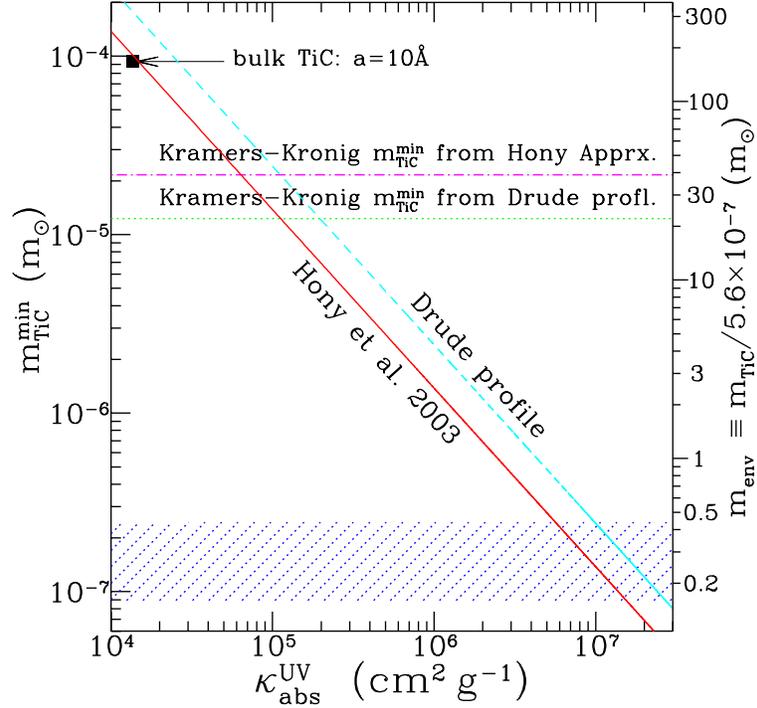}
\end{center}\vspace*{-1em}
\caption{
        \label{fig:mtic}
        \footnotesize
        The TiC mass required to account for 
        the observed 21$\mum$ emission feature
        as a function of the UV/visible mass absorption
        coefficient $\kabsuv$ using the Hony et al.\ (2003)
        hypothetical formula (see Eq.[\ref{eq:kabshony}])
        or an optimized Drude profile (see \S\ref{sec:kk}).
        We also show the TiC mass derived from models
        using the optical constants of bulk TiC 
        (filled square for grains of $a=10\Angstrom$). 
        The horizontal lines plot 
        the Kramers-Kronig lower limits to the TiC mass inferred
        from the wavelength-integrated absorption cross sections
        based on the Hony et al. (2003) approximation 
        (dot-dashed line) and the optimized Drude profile (dotted line)
        (both with $F=1.5$; see \S\ref{sec:kk} and Eq.[\ref{eq:mtickk}]).
        The shaded area indicates reasonable
        values for the total available TiC dust mass (left axis)
        and the total envelope mass 
        including both gas and dust (right axis).          
        }
\end{figure}

In Figure \ref{fig:qabs} we also plot the Hony et al.\ (2003)
hypothetical mass absorption coefficient with 
$\kabsuv=5.6\times 10^{6}\cm^2\g^{-1}$ .
In comparison with the $\kabs(\lambda)$ values
calculated from the optical constants of bulk TiC, 
the Hony et al.\ (2003) formula greatly enhances 
the ability of TiC dust in absorbing the $\HD$ stellar radiation 
which peaks at $\lambda$$\simali$$0.45\mum$.
In Figure \ref{fig:mtic} we show the minimum TiC dust mass $\mticmin$ 
required to account for the observed 21$\mum$ feature 
as a function of $\kabsuv$. 
It can be seen that 
if $\kabsuv \simgt 5.6\times 10^{6}\cm^2\g^{-1}$, 
the minimum required TiC dust mass 
could be smaller than the maximum available TiC mass, 
implying that the TiC model may be tenable.

\begin{figure}[ht]
\begin{center}
\epsfig{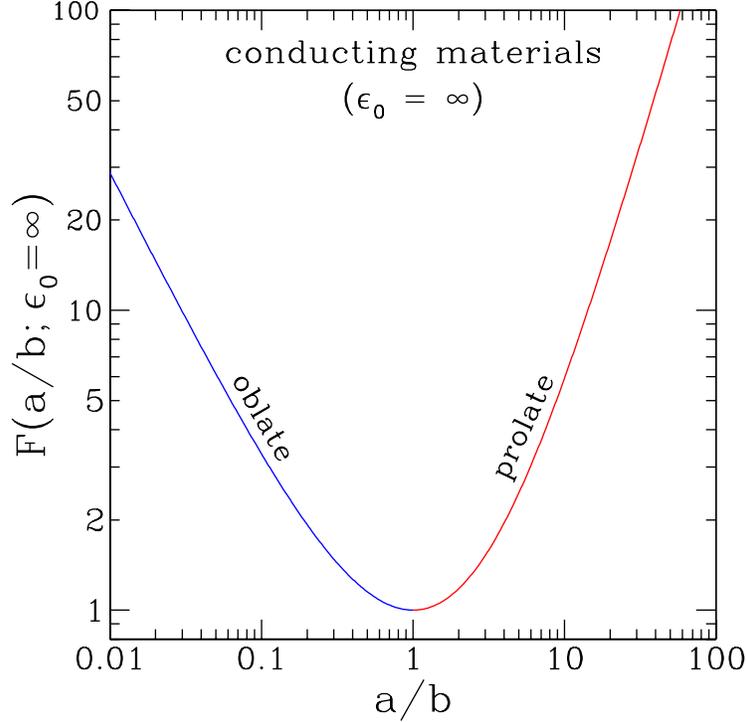}
\end{center}\vspace*{-1em}
\caption{
        \label{fig:dpol}
        \footnotesize
        The $F(\epsilon_0;{\rm shape})$ factor
        as a function of the axial ratio $a/b$
        for highly conducting spheroids 
        (i.e., the static dielectric constant
        $\epsilon_0\rightarrow \infty$).
        For modestly elongated ($a/b\simlt 3$) 
        or flattened ($b/a\simlt 3$) grains,
        the $F$ factor is not expected to 
        significantly deviate from unity. 
        }
\end{figure}

\section{Constraints from the Kramers-Kronig Relations\label{sec:kk}}
Let $\cexttot(\lambda)$ be the total extinction cross sections
of TiC dust at wavelength $\lambda$, 
and $\int_{0}^{\infty} \cexttot(\lambda) d\lambda$ be 
the extinction integrated over the entire wavelength 
range from 0 to $\infty$.
As shown by Purcell (1969), 
the Kramers-Kronig (KK) dispersion relations 
can be used to relate $\int_{0}^{\infty} \cexttot(\lambda) d\lambda$
to the total grain volume $\Vtot$ through
\begin{equation}\label{eq:kk}
\int_{0}^{\infty} \cexttot(\lambda) d\lambda 
= 3 \pi^2 \Vtot F\left(\epsilon_0;{\rm shape}\right) ~~,
\end{equation}
where $F$ is the orientationally-averaged polarizability 
relative to the polarizability of an equal-volume conducting 
sphere (Purcell 1969; Draine 2003).

We can apply Eq.(\ref{eq:kk}) to the $\HD$ circumstellar
envelope to obtain a lower bound on the TiC dust mass $\mtickk$,
taking the envelope to be a vacuum sparsely populated 
by spheroidal TiC grains.\footnote{%
  \label{ftnt:kk}
  This is justified since the total volume of the TiC grains
  ($\simali$$9.0\times10^{27}\cm^3$ 
   for a total mass of $2.2\times10^{-5}\msun$) 
  is negligible compared to the total volume of 
  the dusty circumstellar envelope around $\HD$ 
  ($\simali$$3.1\times10^{51}\cm^3$). Apparently, the envelope 
  also satisfies the prerequisite $|\epsilon-1|\ll1$
  for the KK relations to be applicable
  (where $\epsilon$ is not the dielectric constant of TiC,
  but of the envelope itself).
  } 
Since $\cexttot = \left(\cabstot + \cscatot\right) > \cabstot >0$
(where $\cabstot$ and $\cscatot$ are the total absorption
and scattering cross sections, respectively),
the integration of $\cabstot$ over a finite wavelength range 
represents a lower limit 
to $\int_{0}^{\infty} \cexttot(\lambda) d\lambda$, 
and implies a lower limit to the volume of space which must 
be filled by TiC grains 
and hence a lower limit to the total TiC dust mass
\begin{eqnarray}\label{eq:mtickk}
\nonumber
\mtickk & = & 
\frac{\rhotic}{3\pi^2 F\left(\epsilon_0;{\rm shape}\right)}
\int_{912\Angstrom}^{\infty} \cabstot(\lambda) d\lambda\\
\nonumber
&=& \frac{\mticmin\,\rhotic}{3\pi^2 F} 
\int_{912\Angstrom}^{\infty} \kabs(\lambda) d\lambda\\
&=& \frac{2\rhotic \Eemsn \left(\rmax^2-\rmin^2\right)}
{3\pi^2 \rstar^2\,\ln\left(\rmax/\rmin\right)}
\frac{\int_{912\Angstrom}^{\infty} \kabs(\lambda) d\lambda}
{F \int_{912\Angstrom}^{\infty} \kabs(\lambda)\,\Fstar d\lambda} ~~.
\end{eqnarray}
The dimensionless factor $F$ depends only upon 
the grain shape and the static (zero-frequency) 
dielectric constant $\epsilon_0$ of the grain material 
(Purcell 1969; Draine 2003). 
%The dependence of $F$ on $\epsilon_0$ 
%(up to $\epsilon_0 =300$) for both prolate and oblate 
%spheroids has been calculated 
%by Purcell (1969) and Draine (2003).  
Since TiC is a metallic material, 
%their values may not be applicable.
%We therefore
we calculate the $F$ factors expected for highly
conducting materials (i.e., $\epsilon_0\rightarrow \infty$)
as a function of the axial ratio $a/b$ (where $a$ and $b$ 
are the semiaxis along and perpendicular to the symmetry axis
of the spheroid, respectively).
As shown in Figure \ref{fig:dpol}, $F$ would not appreciably 
exceed unity unless the highly conducting grains 
are highly elongated (for prolates) or flattened (for oblates). 
If they are just modestly elongated 
ot flattened like interstellar grains
($a/b\simali$3 [Greenberg \& Li 1996]
or $b/a\simali$2 [Lee \& Draine 1985]),
we would expect $F\simlt 1.5$. 
Therefore, it is reasonable to adopt 
$F=1.5$ in the following discussions.

It is immediately seen in Eq.(\ref{eq:mtickk}) that
one cannot relax the TiC mass requirement 
just by increasing the UV/visible absorption level;
instead, the Kramers-Kronig lower limit to the total 
TiC dust mass $\mtickk$ is independent of
the absolute level of $\kabs$. 
As discussed in \S\ref{sec:qabs}, the nano-TiC model
appears to be tenable if one adopts the Hony et al.\ (2003) 
$\kabs(\lambda)$ formula with
$\kabsuv \simgt 5.6\times 10^{6}\cm^2\g^{-1}$. 
However, the KK relations (see Eq.[\ref{eq:mtickk}]) 
results in a lower bound of $\mtickk\approx 2.2\times 10^{-5}\msun$ 
to the total TiC dust mass,
exceeding the maximum available TiC mass of
$\mticmax \approx 2.5\times 10^{-7}\msun$ 
(see \S\ref{sec:obs})\footnote{%
  \label{ftnt:ZAMS}
  To be really generous, we can obtain
  an upper limit of $\mticmax \approx 6.1\times 10^{-7}\msun$
  by assuming that all Ti elements in $\HD$ (with a zero-age
  main-sequence mass of $\simali$1.1$\msun$) have condensed 
  in the form of TiC. The TiC model with 
  $\kabsuv \simgt 2.3\times 10^{6}\cm^2\g^{-1}$ 
  (see \S\ref{sec:qabs} and Eq.[\ref{eq:kabshony}])
  would then appear tenable since $\mticmin \simlt \mticmax$. 
  But the KK lower limit on the TiC mass 
  $\mtickk\approx 2.2\times 10^{-5}\msun$
  is far in excess of $\mticmax$
  unless the TiC nanograins are extremely elongated
  ($a/b\simgt 25$) or flattened ($b/a\simgt 80$)
  so that $F\simgt 24$.
  }
by a factor of $\simali$90. Therefore, it appears that 
the nano-TiC model encounters great difficulty in meeting 
the TiC abundance constraint.\footnote{%
  \label{ftnt:a2b2a}
  In order for the TiC dust mass $\mtickk$ derived from the
  KK relations not to exceed the maximum available
  TiC mass $\mticmax$, the TiC nanograins should be extremely
  elongated ($a/b\simgt 44$) for prolates 
  or flattened ($b/a\simgt 220$) for oblates so that $F\simgt59$. 
  But it is hard to imagine how such extremely-shaped 
  TiC nanograins could form and survive in circumstellar
  environments. 
  }

One may argue that the Hony et al.\ (2003) formula 
(Eq.[\ref{eq:kabshony}])
does not make economical use of the TiC absorption. 
A fine tuning on the spectral dependence of $\kabs(\lambda)$
might be able to enhance the grain's ``ability'' to absorb 
stellar radiation, resulting in an increase in
$\int_{912\Angstrom}^{\infty} \kabs(\lambda)\,\Fstar d\lambda$
and a decrease in $\mtickk$ (see Eq.[\ref{eq:mtickk}]).
%In view of the observational fact that the effective 
%temperatures of the 21$\mum$ sources are in the narrow
%range of $5000 \simlt \Teff \simlt 8000\K$,  
We have tried models with a Drude profile-like
UV/visible absorption spectrum\footnote{%
 Spheroidal metallic nanograins are expected
 to have a Drude-profile like resonance if it is
 mainly due to free electrons (Bohren \& Huffman 1983).
 }
$\kabs(\lambda) = \kabsuv\,\left(\gamma\lambda\right)^2/
\left[\left(\lambda^2-\lambda_0^2\right)^2
+ \left(\gamma\lambda\right)^2\right]$
with $\lambda_0=0.44\mum$ and $\gamma=0.35\mum$
(see Fig.\,\ref{fig:qabs})
``{\it tailored}'' to optimize the ``ability'' of 
TiC nanograins to absorb the $\HD$ stellar radiation.
But it is found that models using this $\kabs$ functional 
form require a Kramers-Kronig lower limit of
$\mtickk \approx 1.2\times 10^{-5}\msun$,
still $\simali$50 times larger than the maximum 
available TiC mass.\footnote{%
  A narrower $\kabs(\lambda)$ profile leads to
  a smaller $\mtickk$, provided its peak is fixed
  at $\lambda_0\approx 0.44\mum$. 
  However, even if we adopt an unphysical $\delta$-function
  for $\kabs(\lambda)$, the corresponding KK lower limit 
  $\mtickk \approx 6.8\times 10^{-6}\msun$
  still exceeds $\mticmax$ by a factor of $\simali$27. 
  We note that the grain surface scattering of electrons
  alone would result in a plasma resonance width of
  $\simali$0.1$\mum$ for a 10$\Angstrom$ grain
  a typical Fermi velocity $v_F = 10^{8}\cm\s^{-1}$
  (see Li 2004). 
  }

\section{Discussion}\label{sec:discussion}
In this {\it Letter} we have assumed that the TiC dust
follows the spatial distribution of the bulk carbon
dust (see \S\ref{sec:obs}). 
But this is not critical. Even if we assume that 
all the TiC dust accumulates at the inner edge of 
the envelope, to explain the observed 21$\mum$ feature, 
one still needs $\simali$42\% 
of the total TiC mass derived in \S\S\ref{sec:qabs},\ref{sec:kk}, 
since for the latter the average starlight intensity to which 
the dust is exposed is just $\simali$42\% higher.
Therefore, local density enhancement of TiC 
would not overcome the TiC abundance problem.
High-resolution (FWHM 0.4$^{\prime\prime}$)
imaging at 20.8$\mum$ revealed two bright blobs
at $\simali$1$^{\prime\prime}$ from the star,
but their surface brightness is at most two times 
higher than the average (Kwok, Volk, \& Hrivnak 2002), 
suggesting the insignificance of local density enhancement
of the 21$\mum$ feature carrier.

Von Helden etal.\ (2000) postulated that TiC nanograins
could have formed during the extreme conditions
(i.e., high density and high pressure)
associated with the short, high mass-loss superwind phase
where AGB stars lose the remaining stellar envelope, 
terminating their life on the AGB 
and starting the transition to the PN phase.
However, Kwok et al.\ (2002) argued that
there is no evidence that the 21$\mum$ emission is created 
by a sudden ejection at the end of the AGB,
implying that the necessary physical conditions required to 
form TiC in the ejecta of carbon-rich evolved stars may not be met. 

%In summary, we have shown in this {\it Letter} that 
In summary, although we are lack of experimental knowledge 
of the UV/visible absorption properties of nano TiC, 
a physical argument based on 
the Kramers-Kronig dispersion relations implies that the nano-TiC 
model requires too much Ti to reconcile with the observed 
photospheric Ti abundance.\footnote{% 
  If these grains are coated by a mantle of 
  graphite (see Chigai et al.\ 2003), 
  %(see Chigai, Yamamoto, \& Kozasa 2002),
  the situation would be even worse since 
  the large IR emissivity of graphite
  together with the fact that 
  the increase in grain size may prohibit them from 
  single-photon heating would jointly suppress emission 
  in the 21$\mum$ feature.
  }
Recently, the nano-TiC model was also challenged by
Chigai et al.\ (2003), who found that the abundance ratio 
of Ti to Si needed to reproduce the observed flux ratios 
of the 11.3$\mum$ SiC feature to the 21$\mum$ TiC feature 
of post-AGB stars must be at least 5 times larger than
the solar abundance ratio. Finally, we should note that 
laboratory absorption spectra of TiC nanoclusters 
also show prominent features at other wavelengths which are
not seen in post-AGB stars (e.g. Ti$_{14}$C$_{13}$ also has 
a strong band at $\simali$16$\mum$; van Heijnsbergen et al.\ 1999).

\acknowledgments
I thank D. Arnett, A. Burrows, A.G.W. Cameron, S. Hony, 
D.S. Lauretta and K. Volk for helpful discussions/comments, 
T. Koide and D.W. Lynch for providing me with their optical 
constants of bulk TiC.
I also thank the University of Arizona for the ``Arizona 
Prize Postdoctoral Fellowship in Theoretical Astrophysics''.

\end{document}